\documentclass[11 pt, conference]{IEEEtran}

\usepackage{graphicx}
\usepackage{etoolbox}
\makeatletter
\patchcmd{\@makecaption}
  {\scshape}
  {}
  {}
  {}
\makeatletter
\patchcmd{\@makecaption}
  {\\}
  {.\ }
  {}
  {}
\makeatother
\def\tablename{Table}

\usepackage{hyperref}
\usepackage[utf8]{inputenc}
 
\usepackage{listings}
\usepackage{color}
\usepackage{xcolor}
 
\definecolor{codegreen}{rgb}{0,0.6,0}
\definecolor{codegray}{rgb}{0.5,0.5,0.5}
\definecolor{codepurple}{rgb}{0.58,0,0.82}
\definecolor{backcolour}{rgb}{0.95,0.95,0.92}
 
\lstdefinestyle{mystyle}{
    backgroundcolor=\color{backcolour},   
    commentstyle=\color{codegreen},
    keywordstyle=\color{magenta},
    numberstyle=\tiny\color{codegray},
    stringstyle=\color{codepurple},
    basicstyle=\footnotesize,
    breakatwhitespace=false,         
    breaklines=true,                 
    captionpos=b,                    
    keepspaces=true,                 
    numbers=left,                    
    numbersep=5pt,                  
    showspaces=false,                
    showstringspaces=false,
    showtabs=false,                  
    tabsize=2
}
\lstdefinestyle{ruby}{  
  basicstyle=\ttfamily\footnotesize\color{black},
  commentstyle=\ttfamily\color{red},
  keywordstyle=\ttfamily\color{blue},
  stringstyle=\color{orange},
  breaklines=true,
  breakatwhitespace=false,
  language=Ruby,
  tabsize=2,
  resetmargins=true,
  xleftmargin=0pt,
  frame=single,
  showstringspaces=false
}

\lstdefinestyle{ruby-nocomment}{  
  basicstyle=\ttfamily\footnotesize\color{black},
  commentstyle=\ttfamily\color{black},
  keywordstyle=\ttfamily\color{blue},
  stringstyle=\color{orange},
  breaklines=true,
  breakatwhitespace=false,
  language=Ruby,
  tabsize=2,
  resetmargins=true,
  xleftmargin=0pt,
  frame=single,
  showstringspaces=false
}
 
\ifCLASSINFOpdf
\else
\fi
\begin{document}
%
\title{Reproducible Experiments for Comparing Apache Flink and Apache Spark on Public Clouds}


\IEEEoverridecommandlockouts

\author{\IEEEauthorblockN{Shelan Perera\IEEEauthorrefmark{1}, Ashansa Perera\IEEEauthorrefmark{1}\thanks{\IEEEauthorrefmark{1} These authors contributed equally to this work.}, Kamal Hakimzadeh}
\IEEEauthorblockA{SCS - Software and Computer Systems Department\\
KTH - Royal Institute of Technology\\
Stockholm, Sweden.\\
Email: \{shelanp,ashansa,mahh\}@kth.se
}}


%


\maketitle

\begin{abstract}
Big data processing is a hot topic in today's computer science world. There is a significant demand for analysing big data to satisfy many requirements of many industries. Emergence of the Kappa architecture created a strong requirement for a highly capable and efficient data processing engine. Therefore data processing engines such as Apache Flink and Apache Spark emerged in open source world to fulfill that efficient and high performing data processing requirement. 

There are many available benchmarks to evaluate those two data processing engines. But complex deployment patterns and dependencies make those benchmarks very difficult to reproduce by our own. 

This project has two main goals. They are making few of community accepted benchmarks easily reproducible on cloud and validate the performance claimed by those studies.

\newline
\end{abstract}


%
\IEEEpeerreviewmaketitle

\section{Introduction}
 Today we are generating more data than ever. we are generating nearly 2.5 Quintillion bytes of data per day \cite{HowMuchDataIsGeneratedIBM}. In a world of so much big data the requirement of powerful data processing engines is gaining more and more attention. During the past few years we could observe that there are many data processing engines that emerged. 
 
 Apache Spark \cite{Spark} and Apache Flink \cite{Flink} are such two main open source data processing engines. They were able to accumulate the interest of many data processing stakeholders due to their powerful architectures as well as higher level of performance claims. 
 
 There are many performance reports and articles about Apache Flink  and Apache Spark published by both the industry and the academic institutes. But many studies are not reproducible conveniently even though there are many interested parties to reproduce those results by themselves. There are many valid reasons to reproduce the results because most of the original results are not generic enough to be adopted by others. The requirements may vary from changing the deployment pattern to fine tuning parameters for specific use cases. Data processing engines such as Apache Flink and Apache Spark have such many variables, which can be adjusted so reproducing a completed experiment with different settings is a frequent requirement.
 
 Karamel  \cite{Karamel} a deployment orchestration engine which is developed to design and automate reproducible experiments on cloud and bare-metal environments. It provides a convenient GUI to design reproducible experiments and also a Domain Specific Language (DSL) to declare dependencies and software tools that are required to setup and run the experiment. In this project we are using Karamel to make the distributed experiments reproducible conveniently. Further we are making the experiments and its resources available in Github to facilitate the reproducing those experiments conveniently.

\subsection{Scope}

In this project we are integrating batch processing and stream processing benchmarks to Karamel. We chose Apache Spark and Apache Flink mainly because of their reputation in the data processing world as well as the interest that is being shown by different communities to compare them against known benchmarks. In this report we are evaluating the performance of those two engines with Karamel and discuss how Karamel helps to achieve the goal of reproducibility conveniently.
\newline

\section{Related Work}

We evaluated few benchmarks which are popular among the data processing communities. In this project two different benchmarks were used to design experiments for the batch processing and the stream processing.

\subsection{ Batch Processing}
For batch processing we used Terasort \cite{Terasort} benchmark which was initially developed as a benchmark for evaluating Apache Hadoop Map reduce jobs. This benchmark has been used by many performance validations and competitions, hence considered as one of the most popular applications to benchmark batch processing applications. It was enhanced further to benchmark Apache Flink and Apache Spark by modifying the Map Reduce functionality. 

We integrated the Terasort applications developed by Dongwong Kim \cite{TerasortDongwongKim} for Apache Flink and Apache Spark. He evaluated few data processing frameworks with great details in a clustered environment. His experimental analysis provided a great foundation to start our analysis. We could reuse his experimental code developed for Terasort as it was already analyzed and accepted as a fair analysis to compare the performance by the open source communities.

\subsection{ Stream Processing}
For stream processing we researched two main stream processing benchmarks. They were Intel HiBench Streaming benchmark \cite{HIbenchPaper} and Yahoo Stream benchmark \cite{YahooStreamBenchmark}, which were recently developed to cater the requirement of comparing stream data processing. We evaluated Yahoo Stream benchmark as it includes a more realistic demo of a sample application which simulates an advertising campaign. But We Karamalized HiBench to make it reproducible via Karamel.

\subsection{ Karamel orchestration Engine}
For deployment orchestration we used karamel and reused Chef cookbooks which were developed for clustered deployment of required software with some modifications to adopt them into Karamel. Those cookbooks are building blocks of Karamel,  and Karamel orchestrates the deployment by managing dependencies. We published our cookbooks and Karamel Configurations in Karamel-lab \cite{KaramelLab}  so a user can reuse those cookbook definitions in their experiments.
\newline

\section{Methodology}

This section describes the methodology we used in developing the reproducible experiments. In the first sub section the integration of the Karamel will be briefly described. Next subsections will describe both batch processing and stream processing experiments that were designed to reproduce them on a public cloud.

\subsection{Using Karamel to design the experiments}

Karamel is a web application that can be deployed in your local machine and access its UI through your browser. There are two main logical phases that an experiment designer has to follow when designing an experiment. These phases are logically separated, so parts of the experiments can be developed individually and integrated later into a one experiment. But there is no technical limitation to develop them as a monolithic experiment in one phase.

The first logical phase is the construction of the software-component-deployment and the second logical phase is to develop the experiment that should be run on the deployment. 

The user needs to provide a configuration file which is written in a DSL of Karamel to orchestrate the deployment. The configuration file consists of Karamlized Chef cookbooks and recipes, deployment platform (such as EC2 or Google Cloud) and Node counts. Listing 1 shows a simple hadoop deployment with 3 servers on EC2. Line number 9 shows the cookbook which is used in deployment and their recipes are listed under \(recipes:\) sections, which begins with line number 16 and 23.

\scalebox{.8}{
\begin{lstlisting}[style=ruby,language=Ruby,caption=Sample configuration file for deploying a Hadoop cluster]
name: ApacheHadoopDeployment
#cloud provider
ec2:
    type: m3.medium
    region: eu-west-1
#cookbooks and recipies used in deployment
cookbooks:                                                                      
  hadoop: 
    github: "hopshadoop/apache-hadoop-chef"
    branch: "master"
    
groups: 
  namenodes:
#No of Name nodes in the deployment
    size: 1
    recipes: 
        - hadoop::nn
        - hadoop::rm
        - hadoop::jhs 
  datanodes:
#No of Data nodes in the deployment
    size: 2
    recipes: 
        - hadoop::dn
        - hadoop::nm



\end{lstlisting}
}

This eliminates the burden of deployment and provides a very convenient way to deploy necessary clusters or software on the cloud with a single click.

The second logical phase can be accomplished with the assistance of  inbuilt experiment designer \cite{KaramelExperimentDesigner} in Karamel. A user can use that experiment designer to develop an experiment or modify an existing experiment which is available in the Github. We designed our experiments using the skeletons developed by experiment designer. Listing 2. shows an extract from the generated cookbook which includes the core logic to run an experiment.

There are two ways that a user can design the execution of the experiment. One such way is, a user can combine both the deployment of clusters and execution of the benchmark in a single configuration file. When a user declares a cluster deployment as a dependency for benchmark execution, Karamel creates an intelligent execution graph considering ( Execution DAG) for those dependencies.

Fig. 1. shows a similar Visual DAG generated inside the Karamel. This is a sample DAG generated for Listing 1. configuration file. Since there are 3 nodes in the system, (1 name node and 2 data nodes) those executions could happen in parallel. Further if they have dependencies they are managed intelligently by Karamel and executed in correct dependent order.

\begin{figure}[ht!]
\centering
\includegraphics[width=88mm]{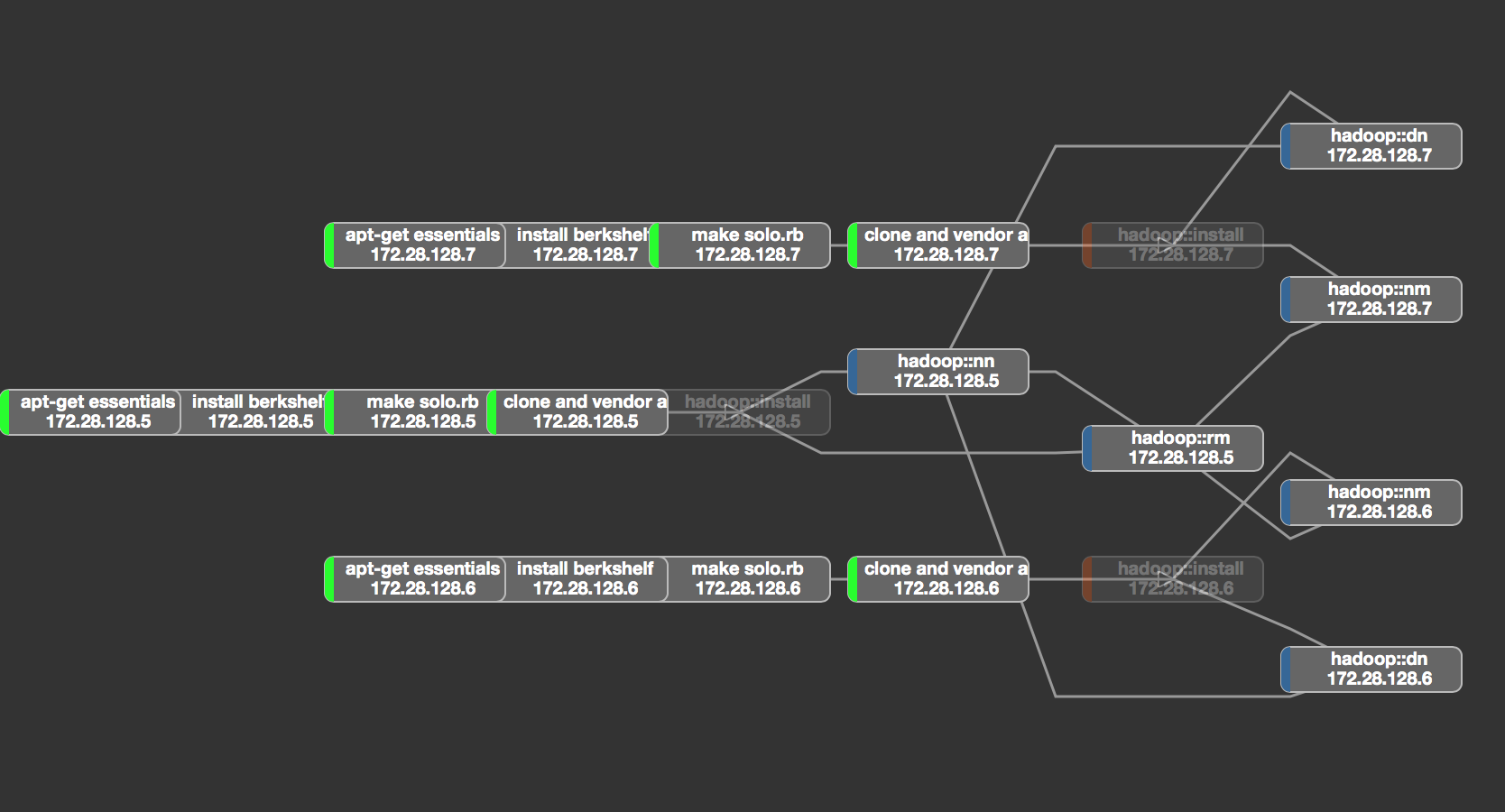}
\label{fig:dag}
\caption{Sample UI populated to change Apache Flink settings}
\end{figure}

\subsection{Designing Batch processing Benchmark Experiment}

Most of the benchmarks have two phases in the benchmark workflow. Benchmark needs to generate sample data to run the core application. Then application can execute its core logic with those input data.
\newline

\subsubsection{Generating Input data for Terasort}

Terasort produces input data using Teragen which internally uses Apache Hadoop map reduce jobs. This can generate large amounts of data that can be used by Terasort application. These generated data is stored in HDFS and used by both Apache Flink and Apache Spark applications.

One line of Teragen has 100 bytes of data. We generated 200GB, 400GB and 600GB of data in size using Teragen with Map Reduce jobs. Teragen data generation is developed as a separate experiment which will run before the experiment.

Listing 2. is the core logic which runs the Teragen application. There is an input parameter [:teragen][:records] (appears in line no 8.) , which can be used to configure the amount of data that needs to be generated. When we launch the experiment we can give the user input from the Karamel UI using generated input fields.
\newline

\scalebox{.8}{
\begin{lstlisting}[style=ruby-nocomment,language=ruby,caption=Part of the Teragen Experiment Cookbook]
script 'run_experiment' do
  cwd "/tmp"
  user node['teragen']['user']
  group node['teragen']['group']
  interpreter "bash"
  code <<-EOM
/srv/hadoop/bin/hadoop fs -rmr /srv
/srv/hadoop-2.4.0/bin/hadoop jar /srv/hadoop-2.4.0/share/hadoop/mapreduce/hadoop-mapreduce-examples-2.4.0.jar teragen #{node[:teragen][:records]} /srv/teragen
  EOM
end
\end{lstlisting}
}

\subsubsection{Running Terasort Experiment}

We developed two experiments for Terasort using Karamel experiment designer similar to Teragen. Both these experiments were ran against the generated data of 200GB, 400GB and 600GB.

\subsection{Designing Streaming Benchmark Experiment}

Streaming benchmarks have a different workflow than the batch processing workflow. In streaming applications data needs to be generated continuously throughout the lifecycle of the experiment. Therefore our experiment was designed to have parallel execution flows for both data generation and data processing by benchmark. As previously stated we made both Yahoo Streaming benchmark and HiBench Streaming benchmark reproducible with Karamel.

\subsubsection{ Yahoo Streaming Benchmark}

Yahoo Streaming Benchmark is designed to test Apache Flink, Apache Spark and Apache Storm \cite{ApacheStorm} streaming performances. This requires additional deployments of Apache Zookeeper \cite{ApacheZookeeper} and Apache Kafka\cite{ApacheKafka} clusters for high throughput stream generation as input data.

This Benchmark simulates a real world advertisement campaign. It
generates ad-campaign data using data generation scripts written in Clojure. All meta data related to ad-campaign is stored in Redis server with the relevant timestamps. Therefore this benchmark updates the Redis database periodically with application level metrics such as latency and throughput. It exposes two parameters to change the behavior of the test. They are:
\begin{itemize}
\item Time duration for Test 
\item No of records sent per second.
\end{itemize}
 By changing those two parameters we could change the level of stress we are imposing at the system. We ran our reproducible benchmarks by changing those parameters and measured how Apache Spark and Apache Flink perform under those conditions. 
 
 We developed Cookbooks to automate the deployment and execution with Karamel. Therefore a user can reproduce the experiments we performed in their cloud environments.
 \newline

\subsubsection{ HiBench Streaming Benchmark}

Intel's Streaming benchmark is based on their HiBench benchmarking project \cite{HiBench}. They have developed it to test the performance of Apache Spark and Apache Storm. There are 7 micro benchmarks such as wordcount, grep and statistics etc. with different complexities and different workloads. Those benchmarks use pre-populated data based on a real world seed data set.

Since it did not have Apache Flink benchmark integrated, we needed to develop an equivalent Apache Flink micro benchmarks for the system. Their benchmark has a complex project structure as they have incorporated streambench as part of their HiBench suite which was developed initially for batch processing.

Similar to the Yahoo benchmark this also uses Apache Kafka and Apache Zookeeper clusters for high throughput stream data generation as the input. They offer different scales of data levels as well as different ways such as push or periodic data generation. In the push mode they try to generate as much as data possible. In the periodic data generation, the amount of data generated per second is throttled by the rate limit.

We also developed a cookbook to manage the lifecycle of the benchmark. This facilitated to execute the benchmark with Karamel as a reproducible experiment conveniently.

\subsection{Monitoring Performance}

In our experiments we collected two different types of performance metrics. They are application level performance metrics and system level performance metrics. For application level metrics we measured metrics such as execution time, latency and throughput. These measurements are experiment and application specific. Therefore we obtained these measurements relative to the experimental setup.

System level metrics can be application independent. Therefore we could use one system level measurement setup across different experiments. We evaluated different system level measurement tools and selected collectl \cite{collectl} due to its light-weightiness and convenient deployment. We developed a comprehensive tool for managing the entire lifecycle of collectl, which we published as collectl-monitoring \cite{CollectlMonitoring} in Github.

Using Collectl-monitoring we generated 4 reports for the system level measurements. They were CPU (CPU frequency and CPU load average per 5 Secs), Memory, Network and Disk utilization. This helps to measure the performance of the system in an application independent manner, without relying on the application specific measurements.

\section{Evaluation}

\subsection{ Batch Processing}

Apache Spark and Apache Flink are memory intensive and process data in the memory without writing to the disks. Therefore we needed to plan our deployments to facilitate the requirements. In batch processing we also needed to store large amounts of data.

We performed few preparation experiments to estimate the storage overhead in our systems. In our experiments we could measure that 3 times of extra space is needed, compared to input data, to store the replicated data as well as temporary and intermediate results. \newline

\subsubsection{Configuration Details}

We chose Amazon Elastic Computing Cloud (EC2) to run our evaluations and we had two types of nodes to be deployed. Some of the nodes act as Master nodes while other nodes act as slaves or worker nodes. Master nodes are processing a very low amount of data while worker nodes undertake the heavy data processing. Therefore, we differentiated them and chose two different types of EC2 instances. We chose m3.xlarge type instances for Master nodes while selecting i2.4xlarge for worker nodes. 

 Table 1. lists the configurations of i2.4xlarge instance and m3.xlarge instance. We used EC2 spot instances \cite{EC2SpotInstances} as we can provide the spot request price limit in the Karamel configuration file, which became a cheaper option for renting EC2 instances.\newline

\begin{table}[!ht]
\renewcommand{\arraystretch}{1.5}
\centering
\begin{tabular}{|c |c |c |} 
 \hline
 & Master (m3.xlarge)
 & Worker (i2.4xlarge)\\ [.5ex]
 \hline
 CPU (GHz) & 2.6 & 2.5 \\ 
   \hline
 No of vCPUs & 4 & 16 \\
   \hline
 Memory (GB) & 16 & 122 \\ 
   \hline
 Storage :SSD (GB) & 80 & 1600 \\[1ex] 
 \hline
\end{tabular}
\vspace*{3mm}
\caption{Amazon EC2 configuration Details for Batch Processing Clusters}
\label{table:1}
\end{table}

\subsubsection{Deployment of Clusters}

In our setup we had 3 main cluster deployments. They are Apache Hadoop cluster, Apache
Flink cluster and Apache Spark cluster. Apache Hadoop cluster is used for running the map reduce programs for Teragen. Teragen produces input data which Apache Spark and Apache Flink clusters used for running the Terasort programs.

Table II. shows how the clusters are deployed in m3.xlarge machine and i2.4xlarge machine. We used 2 instances of i2.4xlarge  and one instance of m3.xlarge.

\begin{table}[!ht]
\renewcommand{\arraystretch}{1.3}
\centering
\begin{tabular}{|c |c |c | c|} 
 \hline
 Cluster
 & Master (m3.xlarge)
 & Worker (i2.4xlarge)& version\\ [.5ex]
 \hline
 
 Hadoop & Name Node & Node Manager & 2.4 \\
        & Resource Manager &  Data Node & \\
        \hline
 Spark  & Spark Master & Spark Worker & 1.3 \\ 
 \hline
 Flink & Job Manager & Task Manager & 0.9 \\[.1ex] 
 \hline
\end{tabular}
\vspace*{3mm}
\caption{Cluster deployments for batch processing experiment}
\label{table:2}
\end{table}

We constructed a Karamle definition and loaded into the Karamel engine to deploy all the clusters in a single execution.\newline

\subsubsection{Executing Terasort Experiment}

After deploying the clusters using Karamel We ran Apache Spark's Terasort experiment and Apache Flink's Terasort experiments separately. After we launched the experiments, we started collectl-monitor to collect system level metrics for the running tests. Once the tests finished its execution, we stopped the collectl-monitor and generated the system level performance reports.

In our deployment we used default configurations and only changed the following parameters to maximize the utilization of memory.

\begin{itemize}
\item Apache Spark

\(spark.executor.memory = 100GB\newline
spark.driver.memory = 8GB\)

\item Apache Flink

\(jobmanager.heap.mb = 8GB\newline
taskmanager.heap.mb = 100GB\)

\end{itemize}

These parameters were changed from Karamel configuration file. It is very convenient for a user to reproduce the results with different memory limits. Listing 3. shows how those parameters can be adjusted using Karamel Configuration file. 

\begin{lstlisting}[style=ruby,language=ruby,caption= Memory settings in Karamel Configuration file]
spark:
    driver_memory: 8192m
    executor_memory: 100g
    user: ubuntu
flink:
    jobmanager:
      heap_mbs: '8192'
    user: ubuntu
    taskmanager:
      heap_mbs: '102400'
\end{lstlisting}

Fig. 2 and Fig. 3 show how those memory parameters for Apache Spark and Apache Flink appear in Karamel configuration UI. This suggests the Karamel's convenience for different stakeholders with different requirements. Anyone who prefers automating everything can benefit from the comprehensive DSL based configuration file. Further, a user who wants to run the experiment adjusting few parameters can change those parameters through Karamel web application UI without knowing Karamel DSL and its configurations.

\begin{figure}[ht!]
\centering
\includegraphics[width=50mm]{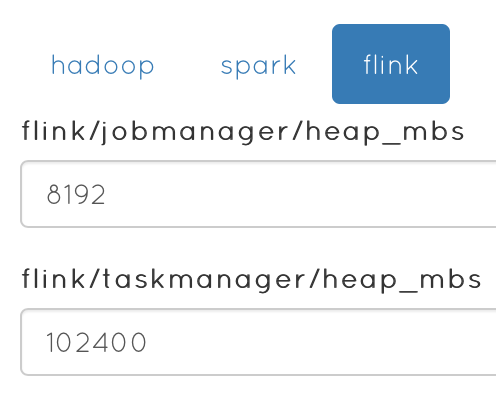}
\label{fig:flink}
\caption{Sample UI populated to change Apache Flink settings}
\end{figure}

\begin{figure}[ht!]
\centering
\includegraphics[width=50mm]{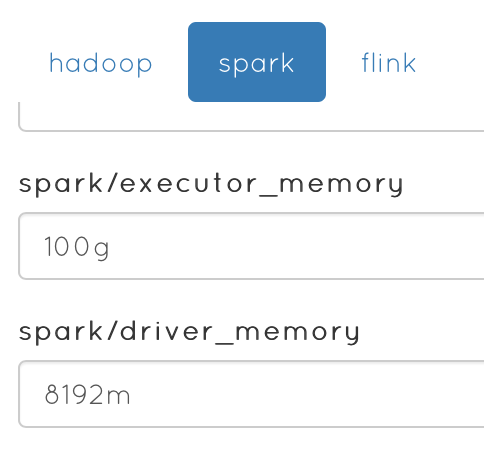}
\label{fig:spark}
\caption{Sample UI populated to change Apache Spark settings}
\end{figure}

Further, our experiments were designed to log the application level details and they were collected to draw the application level comparisons.After running Apache Spark's Terasort and Apache Flink's Terasort we plotted the execution times on a graph to compare the performances.

Fig. 4. compares the execution times logged at application level in the experiments. Those results were obtained at 3 different input levels of 200GB, 400GB and 600GB.

\begin{figure}[ht!]
\centering
\includegraphics[width=80mm]{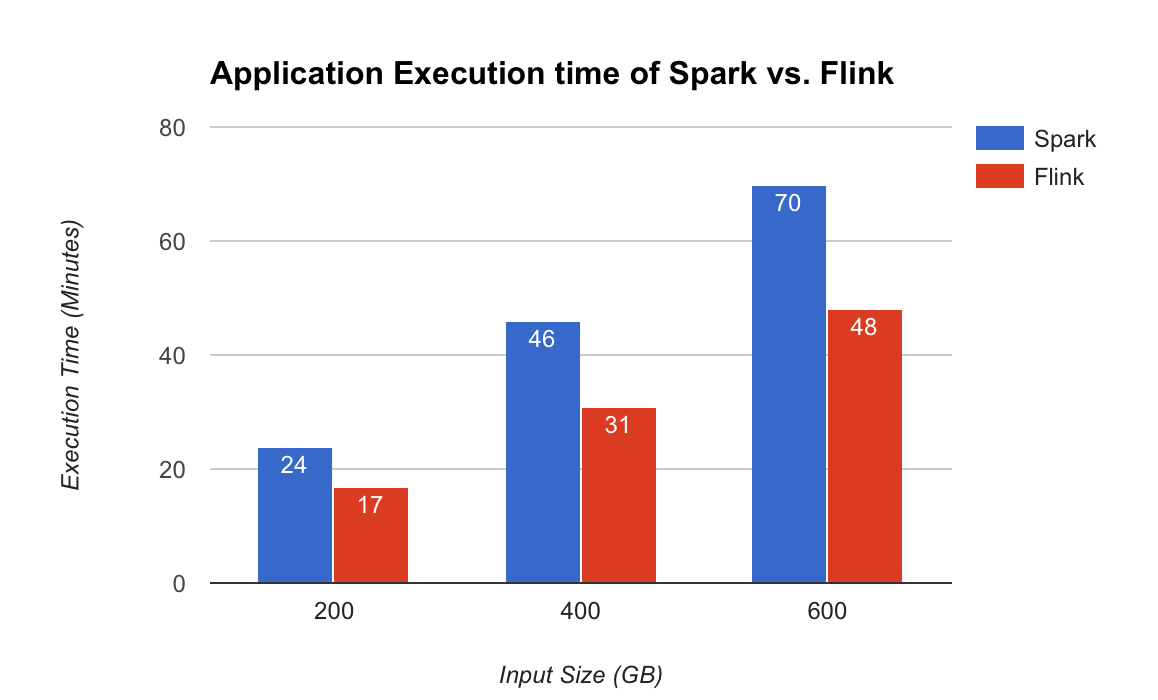}
\label{fig:comparisonofbatch}
\caption{ Execution time comparison for different input sizes}
\end{figure}

We can observe that for all 3 input sizes Flink has recorded less execution time. Therefore Apache Flink has out performed Apache Spark in Terasort application for all 3 workloads we tested. On average Flink was 1.5 times faster than Spark for Terasort application.

Apache Flink is performing better than Apache Spark due to its pipelined execution. This facilitates Flink engine to execute different stages concurrently while overlapping some of them.

For the better explorations we also obtained system level metrics using collectl-monitor and the following graphs illustrates the impact of pipelined execution of Apache Flink.

Fig. 5 and Fig. 6 illustrates the CPU utilization of the two systems.

\begin{figure}[ht!]
\centering
\includegraphics[width=80mm]{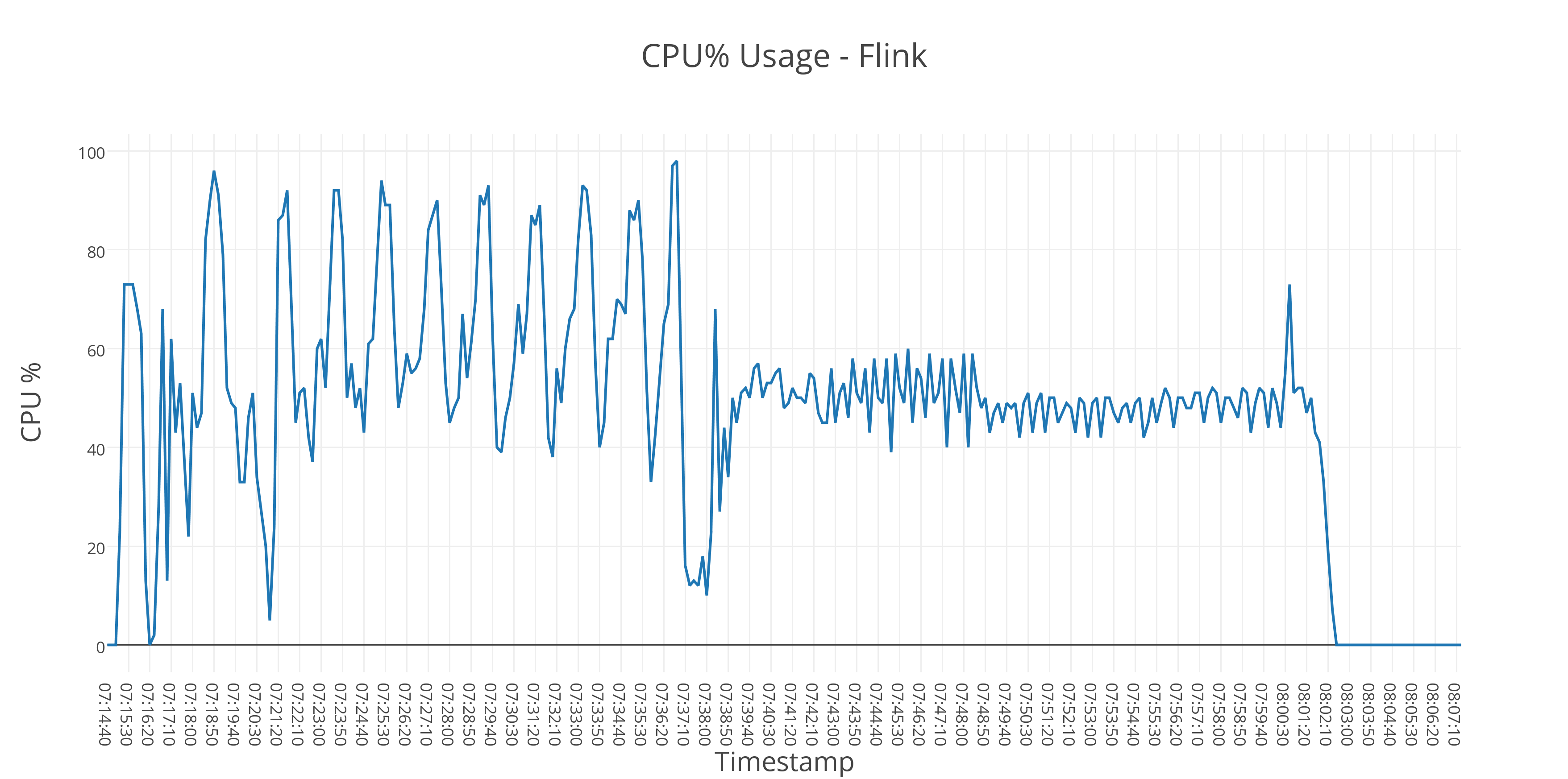}
\label{fig:comparisonofstream}
\caption{CPU utilization of Apache Flink in Batch processing }
\end{figure}

\begin{figure}[ht!]
\centering
\includegraphics[width=80mm]{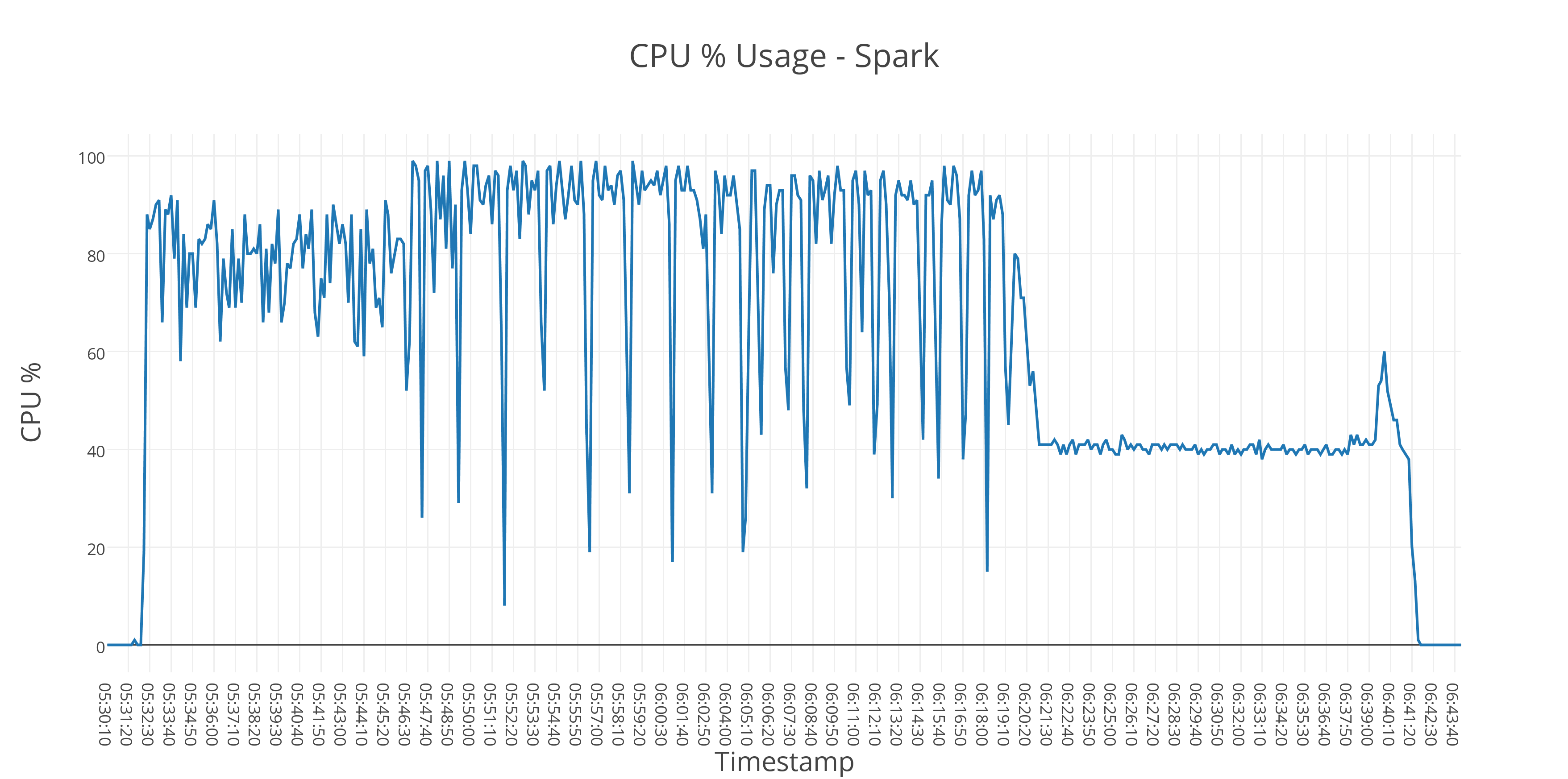}
\label{fig:comparisonofstream}
\caption{CPU utilization of Apache Spark in Batch processing }
\end{figure}

In Fig. 6 We can observe that Apache Spark is reaching almost 100 \% during the execution. But Apache Flink  executed the same load with less utilization as we can observe in Fig. 5. This is a consequence of the pipelined execution as Apache Flink stress the system smoothly while Spark is having significant resource utilization at certain stages.

Fig. 7 and Fig 8  illustrate the Memory utilization of the two systems.

\begin{figure}[ht!]
\centering
\includegraphics[width=80mm]{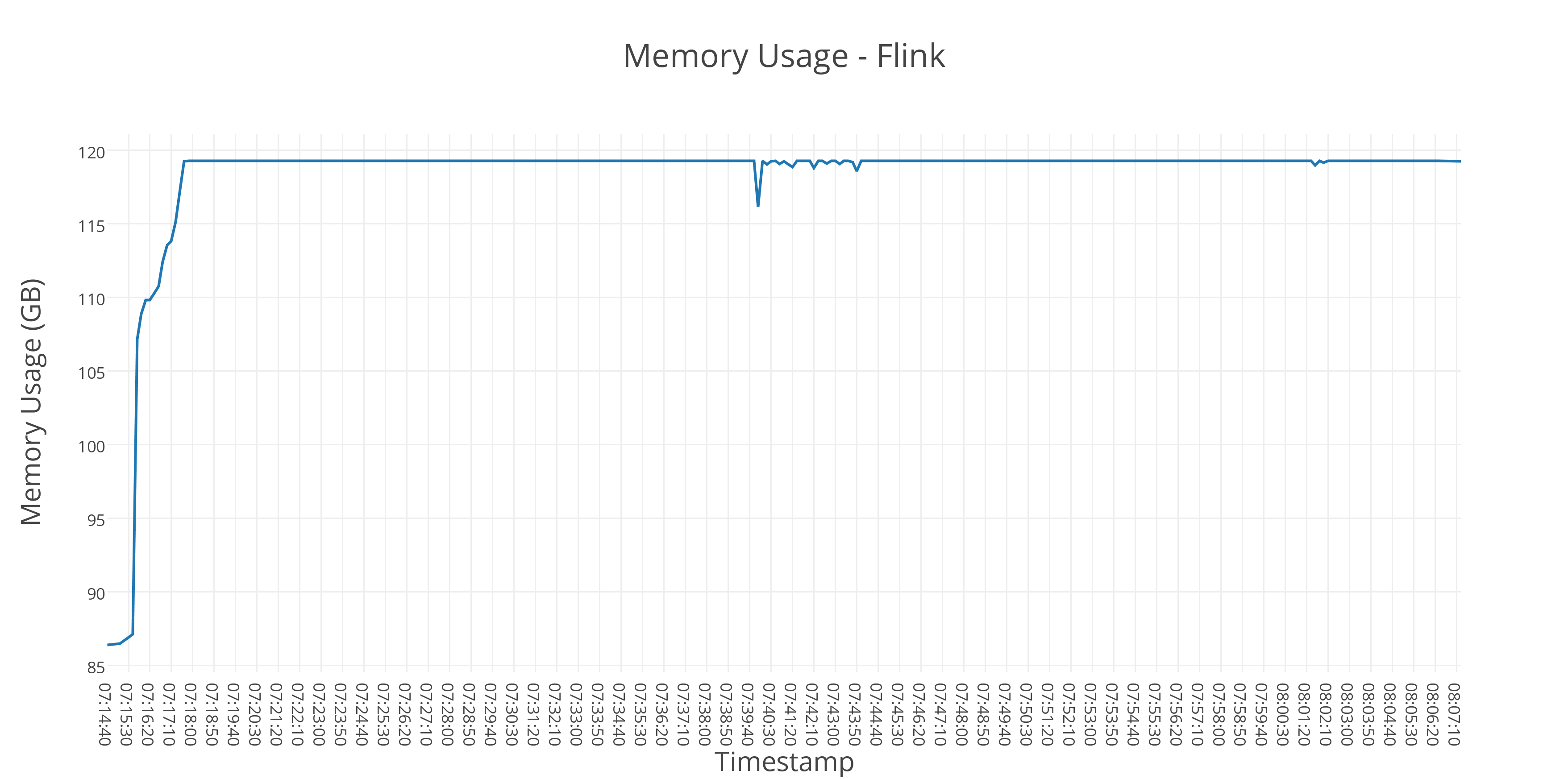}
\label{fig:comparisonofstream}
\caption{Memory utilization of Apache Flink in Batch processing }
\end{figure}

\begin{figure}[ht!]
\centering
\includegraphics[width=80mm]{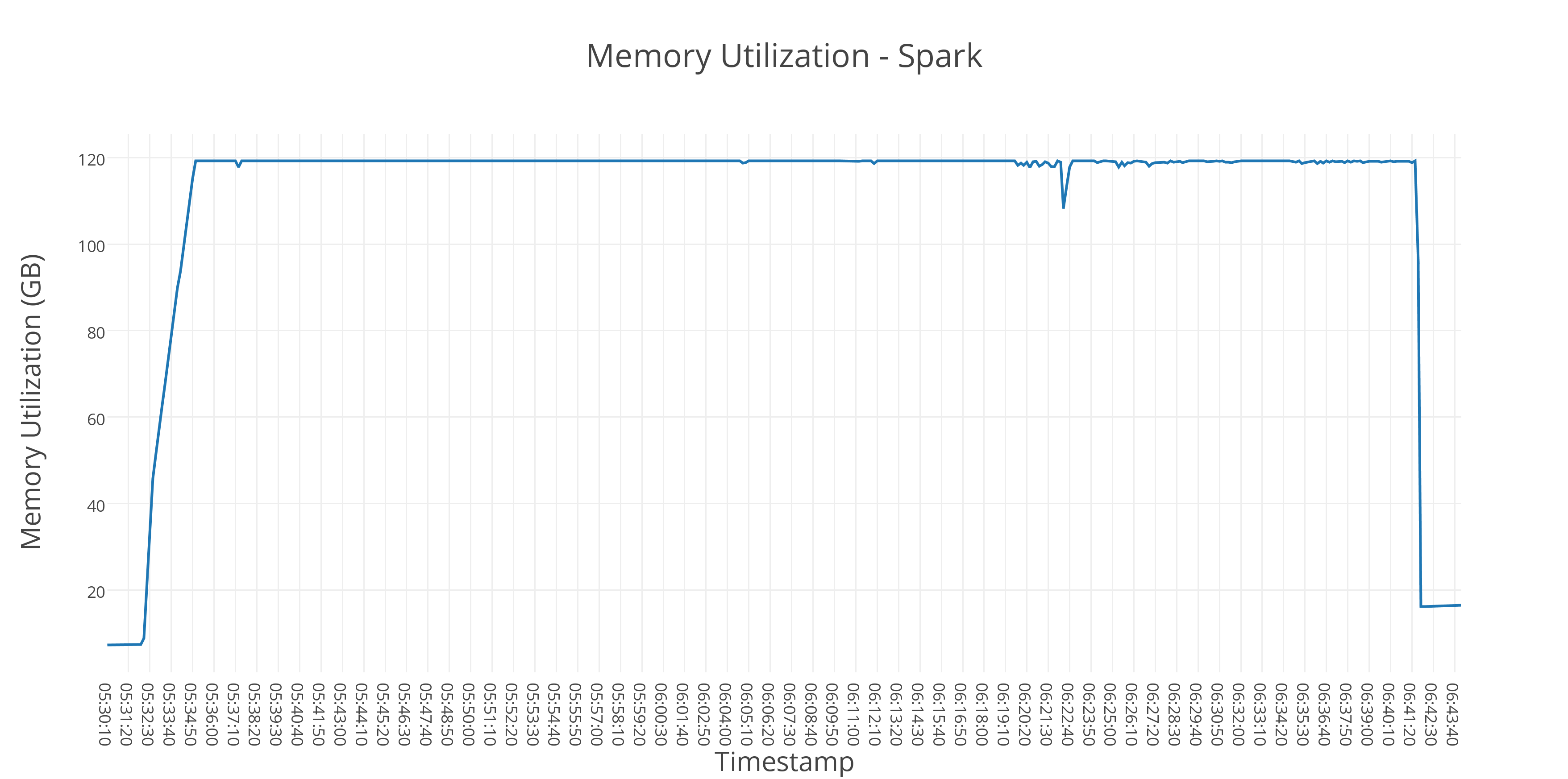}
\label{fig:comparisonofstream}
\caption{Memory utilization of Apache Spark in Batch processing }
\end{figure}

In Fig. 7 and Fig 8 we can observe that both systems utilized all of the memory provided. This is because they are trying to optimize their executions by processing as much as data in the memory without writing to the disks.

Fig. 9. and Fig. 10. illustrate the Disk utilization while Fig 11. and Fig 12. illustrate the Network utilization.

\begin{figure}[ht!]
\centering
\includegraphics[width=80mm]{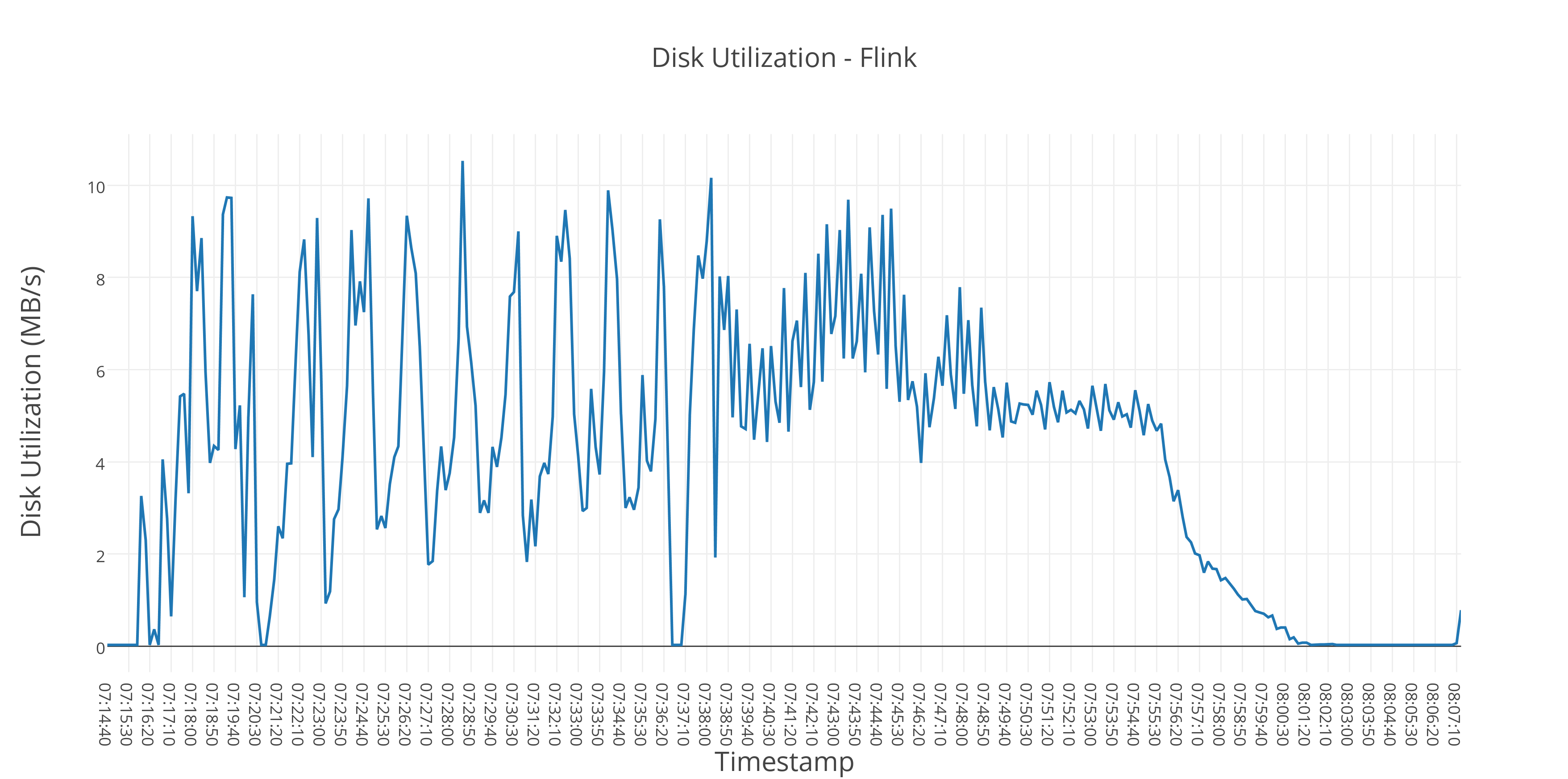}
\label{fig:comparisonofstream}
\caption{Disk utilization of Apache Flink in Batch processing }
\end{figure}

\begin{figure}[ht!]
\centering
\includegraphics[width=80mm]{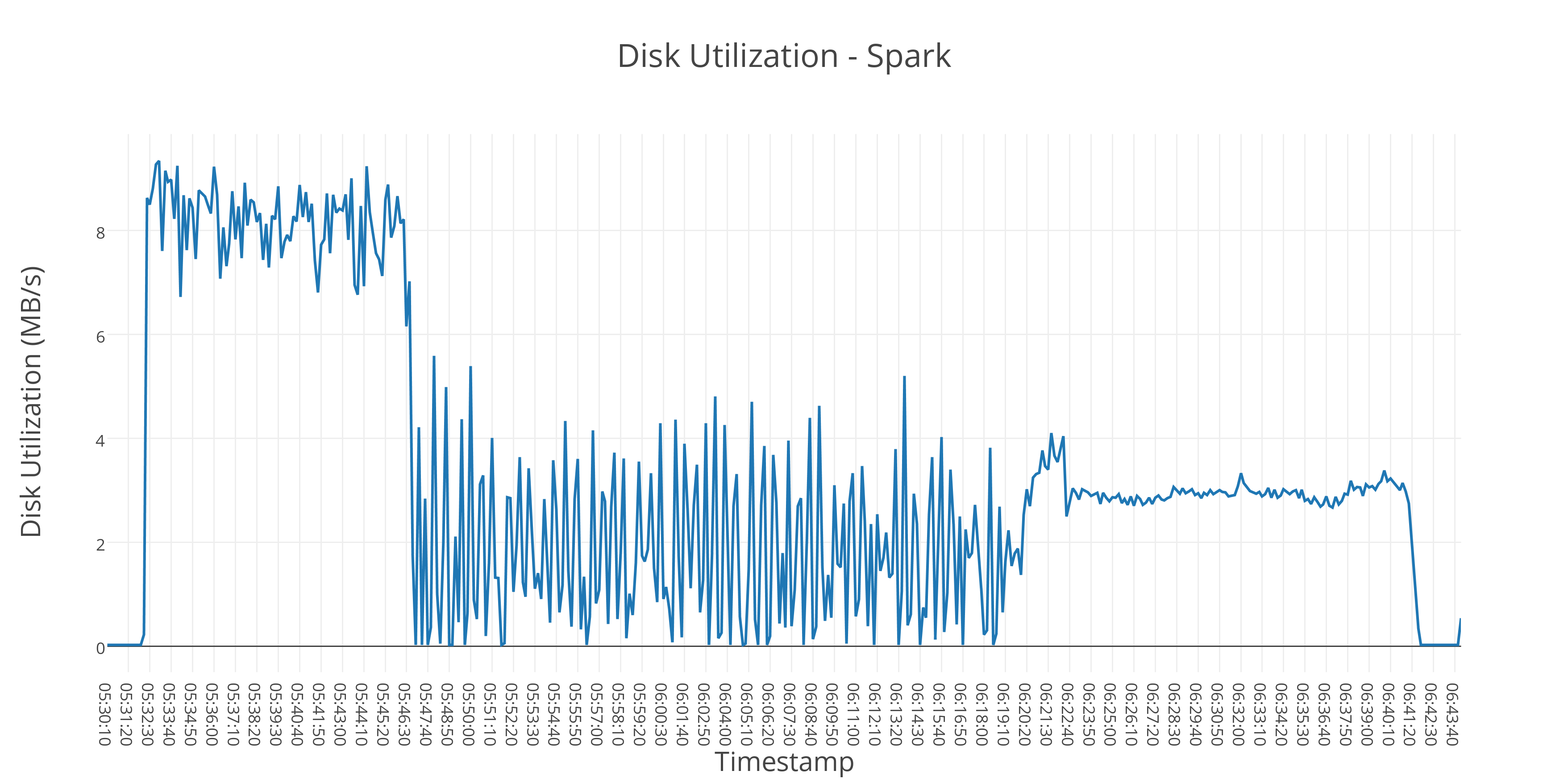}
\label{fig:comparisonofstream}
\caption{Disk utilization of Apache Spark in Batch processing }
\end{figure}

\begin{figure}[ht!]
\centering
\includegraphics[width=80mm]{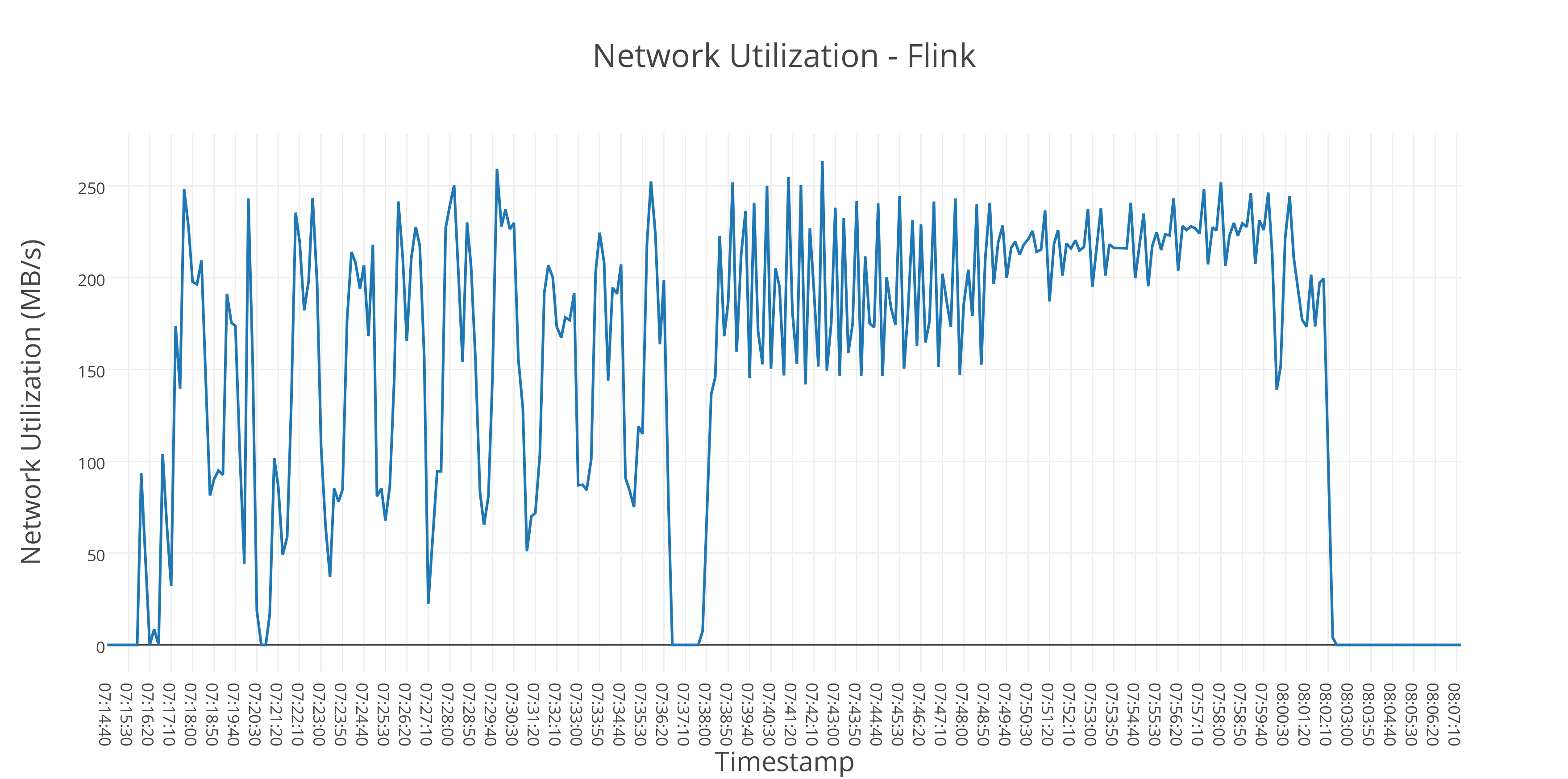}
\label{fig:comparisonofstream}
\caption{Network utilization of Apache Flink in Batch processing }
\end{figure}

\begin{figure}[ht!]
\centering
\includegraphics[width=80mm]{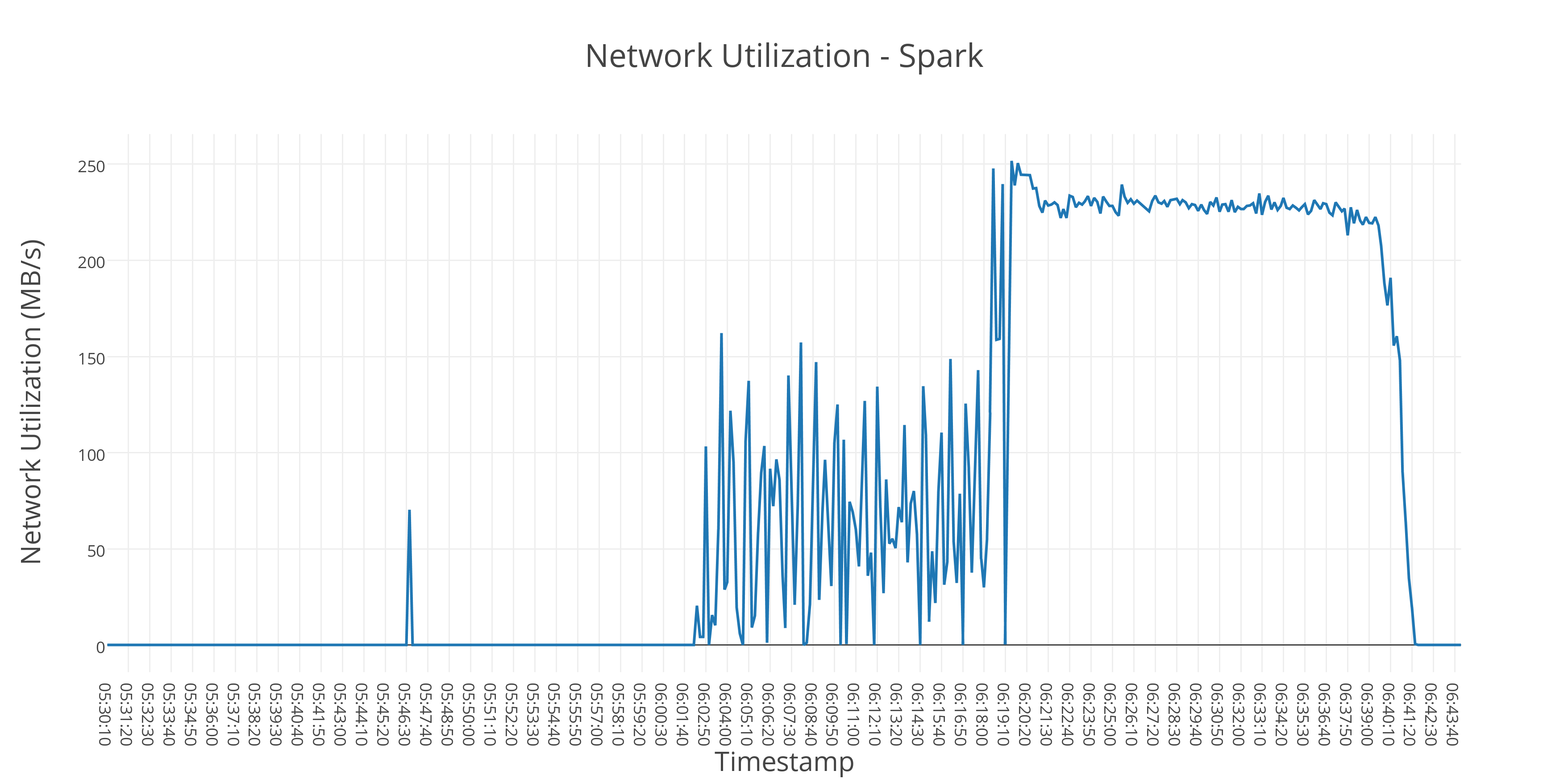}
\label{fig:comparisonofstream}
\caption{Network utilization of Apache Spark in Batch processing }
\end{figure}

When we compare the disk utilization in the Fig. 9. and Fig. 10 , we can observe that Apache Spark clearly shows 3 different phases while Apache Flink shows a similar disk utilization behavior through out. Similarly When we compare the network utilization in Fig. 11 and Fig. 12, we can identify that Apache Spark is having a very insignificant network utilization during the first phase. Since Apache Flink has pipelined execution, it initiates network transfer from beginning.

\subsection{ Stream Processing}

In the stream processing we chose Yahoo Stream benchmarking suite to obtain the comparison results for Apache Flink and Apache Spark. We chose m3.xlarge instances for the low processing nodes similar to our batch processing experiments. But for the high processing node we selected r3.2xlarge instances, as they are optimized for memory intensive applications. i2 instances, which we used for batch processing provided both storage and memory optimizations. But for our streaming application there was a very less storage utilization. 

Table III. shows the details of the configurations we used for streaming applications.

\begin{table}[!ht]
\renewcommand{\arraystretch}{1.5}
\centering
\begin{tabular}{|c |c |c |} 
 \hline
 & Master (m3.xlarge)
 & Worker (r3.2xlarge)\\ [.5ex]
 \hline
 CPU (GHz) & 2.6 & 2.5 \\ 
   \hline
 No of vCPUs & 4 & 8 \\
   \hline
 Memory (GB) & 16 & 61 \\ 
   \hline
 Storage :SSD (GB) & 80 & 160 \\[1ex] 
 \hline
\end{tabular}
\vspace*{3mm}
\caption{Amazon EC2 configuration Details for Streaming clusters}
\label{table:3}
\end{table}

\subsubsection{Deployment of Clusters}

In our setup we had 3 main cluster deployments as similar to batch processing deployments. They were Apache Hadoop cluster, Apache Flink cluster and Apache Spark cluster. Additionally Apache Zookeeper and Apache Kafka were deployed for generating input data. We deployed them in the high processing node without making them fully distributed, to match the configurations of the Yahoo benchmark.This deployment facilitated minimal modifications to the benchmark execution scripts.

Table IV. shows how the clusters are deployed in m3.xlarge machine and r3.2xlarge machine.In this deployment we used 1 instance of r3.xlarge  and one instance of m3.xlarge.

\begin{table}[!ht]
\renewcommand{\arraystretch}{1.3}
\centering
\begin{tabular}{|c |c |c | c|} 
 \hline
 Cluster
 & Master (m3.xlarge)
 & Worker (i2.4xlarge)& version\\ [.5ex]
 \hline
 
 Hadoop & Name Node & Node Manager & 2.7.1 \\
        & Resource Manager &  Data Node & \\
        \hline
 Spark  & Spark Master & Spark Worker & 1.5.1 \\ 
 \hline
 Flink & Job Manager & Task Manager & 0.10.1 \\[.1ex] 
 \hline
 Zookeeper & - & Zookeeper & 3.3.6 \\[.1ex] 
 \hline
 Kafka & - & Kafka brokers & 0.8.2.1 \\[.1ex] 
 \hline
 Redis & - & Redis database & 3.0.5 \\[.1ex] 
 \hline
 
\end{tabular}
\vspace*{3mm}
\caption{Deployment versions of clusters}
\label{table:4}
\end{table}

Similar to the approach that we followed in the batch processing experiment we deployed the clusters using Karamel with a Karamel configuration file we developed. These scripts are hosted in Karamel lab \cite{KaramelLab} for a user to reproduce easily in a cloud environment.

We ran the benchmarks with the assistance from Karamel for different streaming rates. We configured event rates, starting from 1000 events per seconds up to 10000 events per second by increasing the rate with 1000 steps.

Fig. 13 shows the behavior of Apache Spark and Apache Flink with different streaming events rates. This graph was plotted by aggregating the latencies of different stream executions and obtaining 99th percentile of the data respectively.

\begin{figure}[ht!]
\centering
\includegraphics[width=80mm]{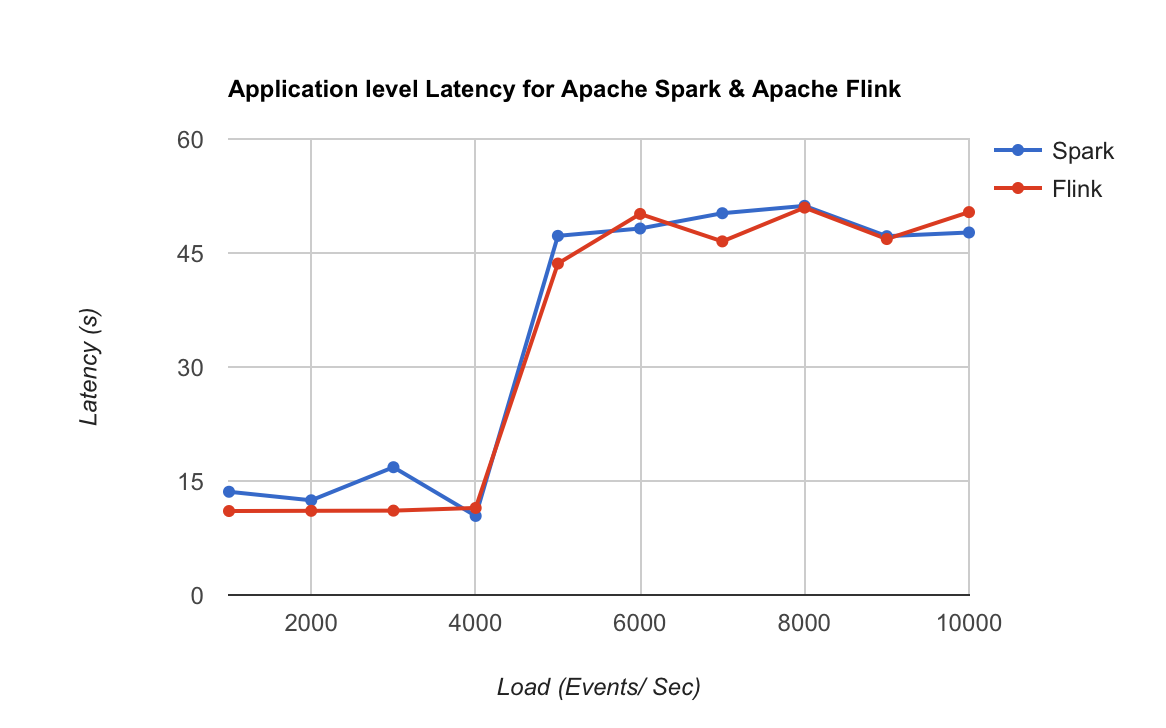}
\label{fig:comparisonofstream}
\caption{Latency comparison for different event rates }
\end{figure}

We could observe that both Apache Flink and Apache Spark performed similarly under different loads. But we could observe that Apache Flink demonstrated a slightly better performance at lower event rates such as 1000, 2000 and 3000 events per second. 

The reason for this slight improvement might have caused by Apache Flink's architecture. Apache Flink supports native stream execution but Apache Spark is approximating streaming by using a micro batch model. In Apache Spark they execute small micro batches per event window and apply operations.

There was a significant change in the latencies after 4000 events /sec input rate. This happened because of the Apache Kafka setup that was configured with the Yahoo Stream bench consisted only single broker for reducing the complexity. Therefore it could not handle that amount of input rate and generated a back pressure scenario.

In overall both Apache Spark and Apache Flink behaved in a similar manner for different input stream loads. In Yahoo benchmark the latencies reported, were the time between last event was emitted to kafka for a particular campaign window and when it was written into Redis. Therefore it is not the exact latency for Apache Flink or Apache Spark, but a relative measurement to compare the performance relatively.

We also collected system level metrics using Collectl-monitor. Fig. 14 and Fig. 15 shows the CPU utilization while running the benchmark. We could observe similar CPU behavior in both of the systems. Since this CPU utilization also includes the data initialization (seed data generation) and post processing (computing the latencies from Redis timestamps) of the Yahoo streaming benchmark, there are corresponding spikes for those stages in the system.

Fig. 16 and Fig. 17 has less memory utilization than the batch processing as we can see in Fig. 7 and Fig. 8. But in streaming Apache Flink consumed less amount of memory than Apache Spark. We can observe that Apache Flink consumed less than 14GB of memory while Apache Spark consumed around 18GB during its execution.

\begin{figure}[ht!]
\centering
\includegraphics[width=80mm]{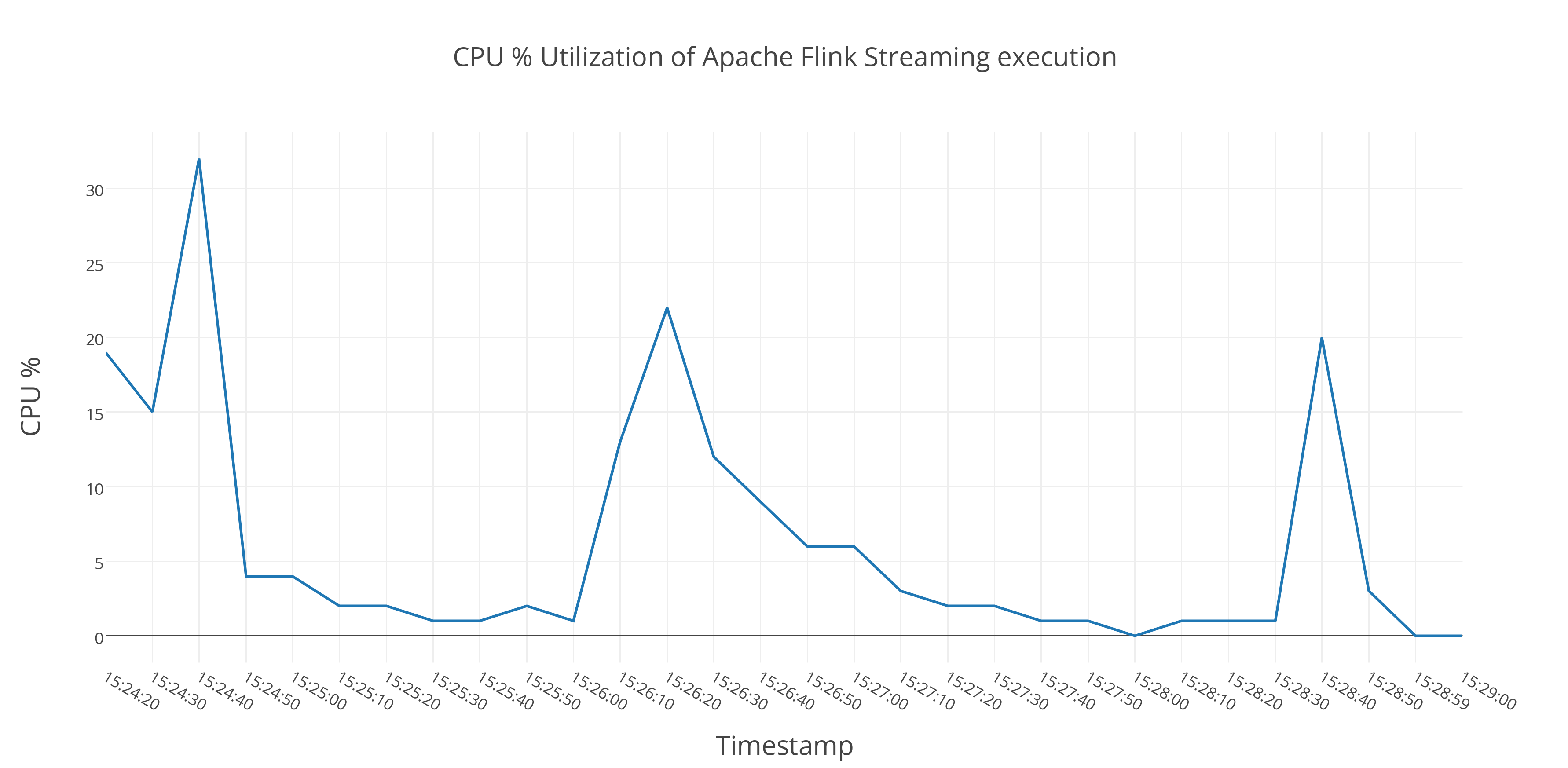}
\label{fig:comparisonofstream}
\caption{CPU utilization of Apache Flink in Stream processing }
\end{figure}

\begin{figure}[ht!]
\centering
\includegraphics[width=80mm]{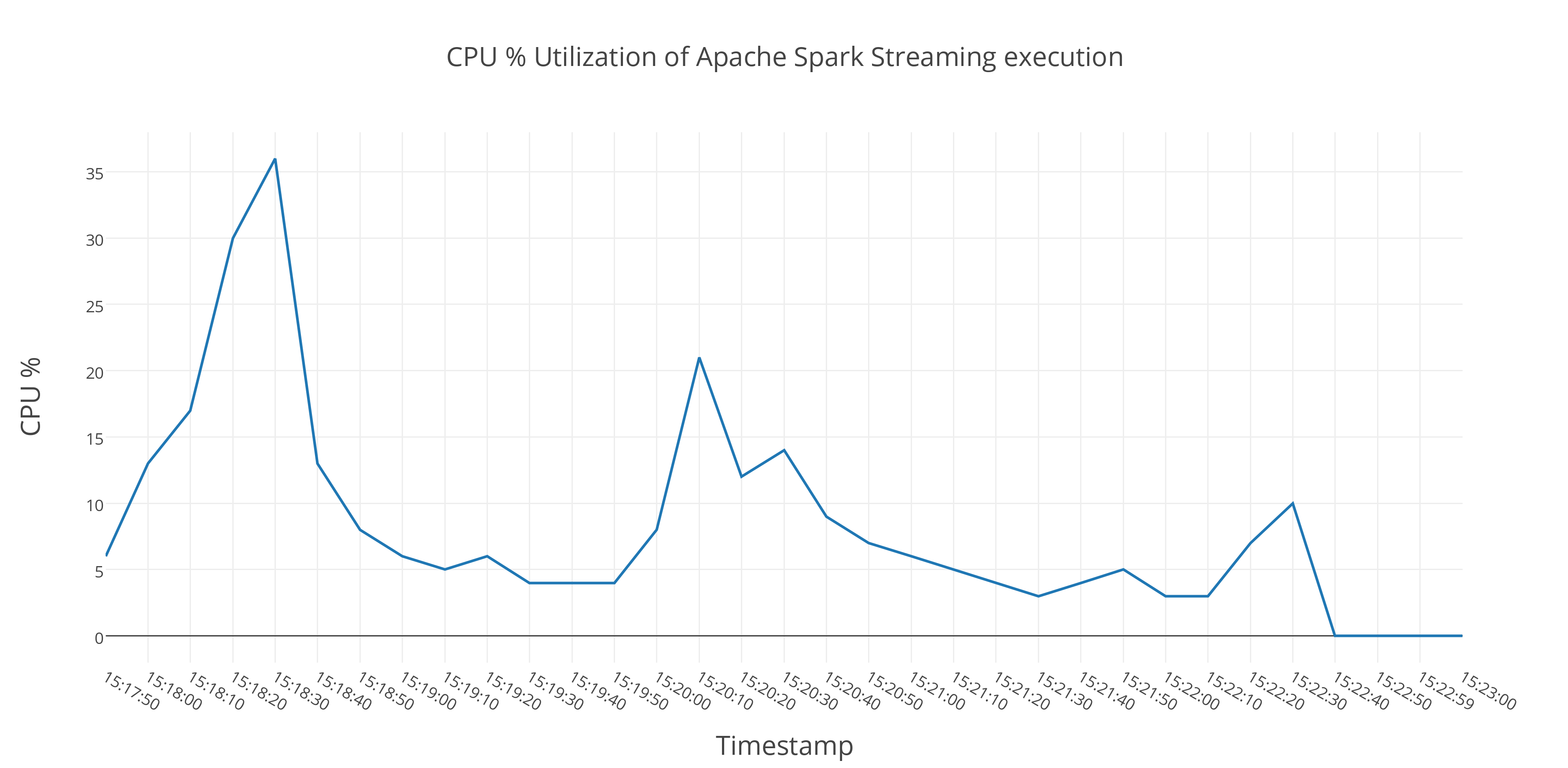}
\label{fig:comparisonofstream-cpu-spark}
\caption{CPU utilization of Apache Spark in Stream processing }
\end{figure}

\begin{figure}[ht!]
\centering
\includegraphics[width=80mm]{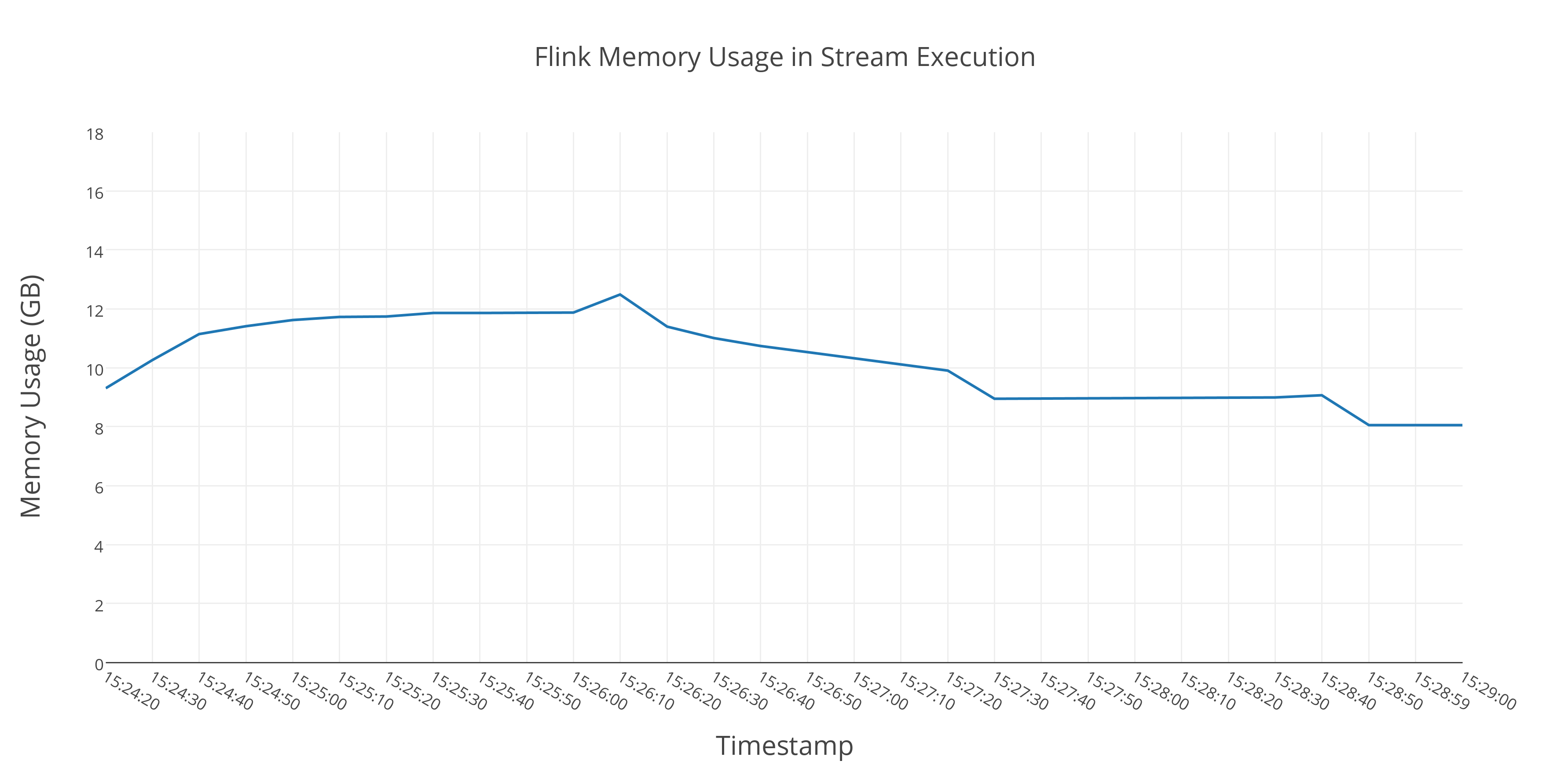}
\label{fig:comparisonofstream}
\caption{Memory utilization of Apache Flink in Stream processing }
\end{figure}

\begin{figure}[ht!]
\centering
\includegraphics[width=80mm]{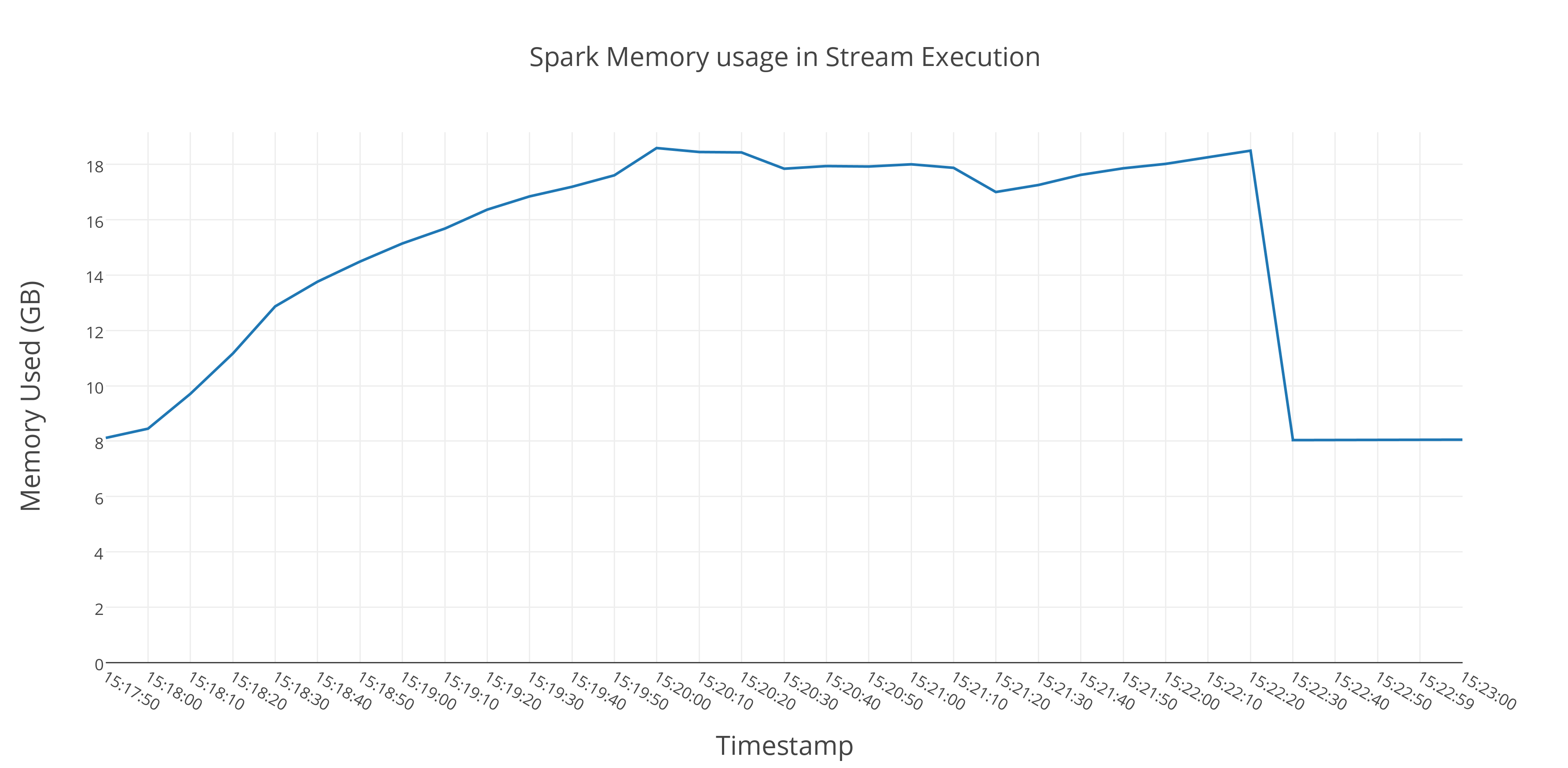}
\label{fig:comparisonofstream}
\caption{Memory utilization of Apache Spark in Stream processing }
\end{figure}

Both application level and system level metrics do not clearly highlight, which data processing engine is performing better. Eventhough, we could observe that Flink was performing slightly better in many occasions than Spark.

\subsection {Reproducibility with Karamel}

One of our main goals of the project is to make the benchmark applications conveniently reproducible. This convenient reproducibility covers two main phases. They are convenient deployment at deployment of the clusters, and convenient execution of the benchmark applications.\newline

\subsubsection{ Convenient deployment of clusters}

After starting the Karamel web application a user can reuse the Karamel configuration file for deploying the Apache Flink, Apache Hadoop, Apache Spark clusters. For stream processing applications we used two more deployments of Apache Kafka and Apache Zookeeper. Both Karamel configurations are available at Karamel Lab \cite{KaramelLab} for a user to load into their own Karamel web application and deploy in their own cloud provider. 

There is a flexibility for user to change the parameters through configuration file as mentioned in Listing 3. or to change them through the Karamel UI as shown in Fig 2. and Fig 3. Similarly no of nodes that needs to be deployed can be altered by changing the \(size\) parameter (eg: listing 1. line no. 15 or
line no. 22). So a user can recreate clusters very conveniently without undergoing the painful process of configuring the clusters.

\subsubsection{ Convenient execution of the benchmark}

Similar to deploying the clusters a user can execute the benchmark on the deployed cluster. Similar to the cluster deployment, Parameters that are exposed
can be changed either through UI parameter or using the Karamel configuration file. This allows a user to run the same experiment repeatedly many times with different settings. 

Therefore, these features of Karamel provides a smooth end to end workflow for users to reproduce their experiments many times with much less overhead.

\section{Conclusion}

In this project we developed reproducible experiments for Apache Spark and Apache Flink in the cloud. Therefore another user who is interested in reproducing the same experiments can reproduce in their own cloud environments. Further they can easily change the exposed parameters to deeply analyse different behaviours at different scales.

We could achieve the goal of making the experiments publicly available  by sharing all the reproducible artifacts online in Karamel Lab \cite{KaramelLab}. Further in our experiments we could validate the claims of the few studies such as Dongwong Kim's performance comparison which compared the performance of Apache Flink and Apache Spark. We could obtain similar characteristics and results in our setup therefore strengthening the results found by those initial comparisons.

\section{Future Work}

Intel's HiBench and Yahoo Stream Benchmarks were developed recently. Therefore there are few issues that needs to be considered to evolve them Karamel. The main reason is they were designed to run manually in a deployed cluster not with an automated orchestrated environment like Karamel.

Yahoo's Benchmark experiment needs to be enhanced to run with a fully distributed clusters of Kafka and Zookeeper to facilitate the experiment with very high input rates. 

This complete project can be a basic framework for re-executing Apache Flink and Apache Spark benchmarks on public clouds to run different experiments in future.





%

\nocite{*}
\bibliography{main}
\bibliographystyle{ieeebib}

\end{document}